\documentclass[epj,nopacs]{svjour}
\usepackage{epsfig,array}
\usepackage{multirow,color}
\usepackage{epsf,graphicx}
\usepackage{times,latexsym,euscript,amstext,amssymb,amsbsy,amsmath,amsfonts}
\usepackage{mdwlist}\usepackage{epsfig}
\usepackage{slashbox,multirow}
\usepackage[section] {placeins}
\newcommand{\be}{\begin{eqnarray}}
\newcommand{\ee}{\end{eqnarray}}
\newcommand{\bc}{\begin{center}}
\newcommand{\ec}{\end{center}}
\newcommand{\bea}{\begin{eqnarray}}
\newcommand{\eea}{\end{eqnarray}}

\newcommand{\beq}{\begin{equation}}
\newcommand{\eeq}{\end{equation}}

\def\fun#1#2{\lower3.6pt\vbox{\baselineskip0pt\lineskip.9pt
\ialign{$\mathsurround=0pt#1\hfil##\hfil$\crcr#2\crcr\sim\crcr}}}

\begin{document}

\title{\boldmath Sign ambiguity in the $K\Sigma$ channel}
\titlerunning{Sign ambiguity in the $K\Sigma$ channel}
\author{
A.V.~Anisovich$\,^{1,2}$, E.~Klempt$\,^1$, V.A.~Nikonov$\,^{1,2}$,
A.V.~Sarantsev$\,^{1,2}$,  and U.~Thoma$\,^{1}$ }
\authorrunning{A.V.~Anisovich \it et al.}
\institute{$^1\,$Helmholtz-Institut f\"ur Strahlen- und Kernphysik,
Universit\"at Bonn, Germany\\
$^2\,$Petersburg Nuclear Physics Institute, Gatchina, Russia}

\date{Received: \today / Revised version:}

\abstract{Ambiguities of the signs of $N\to \Sigma K$ coupling
constants are studied in a multichannel partial wave analysis of a
large body of pion and photo-induced reactions. It is shown that the
signs are not free from some ambiguities, and further experimental
data are needed. Data on the reactions $\pi^+p\to \Sigma^+K^+$ and
$\gamma p\to K^+\Sigma^0$ define rather well the isospin $3/2$
contributions to these channels. However the lack of information on
polarization observables for the reactions $\pi^-p\to \Sigma^0K^0$,
$\pi^-p\to \Sigma^-K^+$ and $\gamma p\to K^0\Sigma^+$ does not allow
us to fix uniquely the signs of $N\to \Sigma K$ coupling constants.
As a consequence, also the contributions of nucleon resonances to
these channels remain uncertain.}

\maketitle

\section{Introduction}
The precision and diversity of data on photo-induced reactions off
protons and neutrons studied experimentally has increased rapidly in
the past, and significantly more data are expected in the near
future. The data comprise high-precision differential cross sections
of various reactions and data in which the initial photons and/or
the target nucleons are polarized, and data in which the
polarization of final-state baryons is recorded. Recent reviews of
ideas and results in baryon spectroscopy can be found in
\cite{Klempt:2009pi,Crede:2013kia}. Since then, important steps have
been made by several groups analyzing pion and photo-induced
reactions in coupled-channel frameworks. Here, we remind the reader
of the recent developments.

The Giessen group has pioneered coupled-channel analyses of large
data sets \cite{Feuster:1997pq,Feuster:1998cj,Penner:2002ma,Penner:2002md,%
Shklyar:2004dy,Shklyar:2004ba,Shklyar:2005xg,Shklyar:2006xw,Shklyar:2012js}.
Their most recent paper \cite{Cao:2013psa} focusses on pion and
photo-induced reactions of $\Sigma K$ final states with the aim to
extract the couplings of known resonances to the $K\Sigma$ state.

The Bonn-J\"ulich group analyzed isospin $I=3/2$ $\pi N$ elastic
scattering amplitudes from the GWU/SAID analysis~\cite{Arndt:2006bf}
jointly with data on the $\pi^+ p \to K^+ \Sigma^+$
reaction~\cite{Doring:2010ap}. The analysis was extended to
isospin-1/2 contributions by including all $\pi N$ elastic
scattering amplitudes from \cite{Arndt:2006bf} and the reactions
$\pi N\to N\eta$, $\Lambda K$, and $\Sigma K$ \cite{Ronchen:2012eg}.
A consistent treatment of the three $\pi p\to \Sigma K$ channels
($\pi^+ p \to\Sigma^+ K^+$, $\pi^- p \to\Sigma^- K^+$, and $\pi^- p
\to\Sigma^0 K^0$) was reached in \cite{Ronchen:2012eg} but required
- compared to \cite{Doring:2010ap} - significant changes in the
relative importance of the contributions to the total cross section
from different partial waves. Photoproduction of single pions was
included in a study presented in~\cite{Huang:2011as}. It is shown
that a good description of the data can be achieved.

The Osaka-Tokyo-Argonne group studies baryon resonances in a
dynamical coupled-channels model by fitting a large body of pion and
photo-induced reactions \cite{Matsuyama:2006rp,JuliaDiaz:2007fa,%
Kamano:2008gr,Kamano:2009im,Kamano:2013iva}.

The Bonn-Gatchina group has published recently a comprehensive
analysis of a large body of pion and photo-induced reactions
\cite{Anisovich:2011ye,Anisovich:2011fc,Anisovich:2012ct}. At
present, this is the only group which systematically searched for
new resonances in all partial waves. Mass, width, and partial decay
widths of many resonances - including several new baryon resonances
- were determined; their errors were evaluated by a systematic
variation of the model parameters. The final results can be found in
the latest RPP \cite{Beringer:1900zz}. The new resonances disfavor
\cite{Anisovich:2011su} conventional diquark models in which one
pair of quarks is frozen into a quasi-stable diquark
\cite{Anselmino:1992vg}. The observed pattern of resonances seems to
occupy fully a limited number of  SU(6) multiplets
\cite{Anisovich:2011sv} while other multiplets remain void.
Interestingly, the new resonances can all be grouped naturally into
spin-parity doublets. At present, there is an ongoing discussion if
the occurrence of parity doublets in meson and baryon spectroscopy
evidences a phase transition from broken to restored chiral symmetry
\cite{Glozman:1999tk,Glozman:2007ek,Jaffe:2004ph}.

The Kent group \cite{Shrestha:2012ep} has updated an older analysis
\cite{Manley:1984jz} of $\pi N$ elastic scattering amplitudes and
(low-statistics) bubble chamber data on $\pi N\to N\pi\pi$. The
group confirmed the existence of most resonances reported in
\cite{Anisovich:2011fc}. Some of them had already be seen in their
1983 analysis \cite{Manley:1984jz} even though the Particle Data
Group did not open new entries for these resonances at that time.

Several groups studied particular aspects. The Gent group developed
a Regge-plus-resonance (RPR) model~\cite{Corthals:2005ce} in which
the background is deduced from the high-energy Regge-trajec\-tory
exchange in the t-channel to which a few resonances are added. The
coupled-channel model of the Groningen
group~\cite{Usov:2005wy,Shyam:2008fr} was later extended at Giessen
\cite{Shyam:2009za}. The fit uses established resonances and derives
decay coupling constants which are compared to SU(3) relations.

In the present paper we present the results of a systematic
investigation of reactions with $K\Sigma$ final states within the
Bonn-Gatchina partial wave analysis. All results were obtained in a
combined analysis of data on $\pi N$, $\eta N$, $K\Lambda$,
$K\Sigma$, $\pi\pi N$ and $\pi\eta N$ final states
\cite{Anisovich:2011ye,Anisovich:2011fc,Anisovich:2012ct} including
recent measurements from the CBELSA/TAPS \cite{Schmieden:2012zz} and
MAMI-C \cite{Jude:2013jzs} collaborations. In total, 31.180 data
points from two-body reactions are used which are described with a
$\chi^2$ of 48.710, or $\chi^2/N_{\rm F}=1.6$. The fit is further
constrained by a fraction of the events ($\approx$ 500.000) from
three-body final states which are included in an event-based
likelihood fit.

\section{\boldmath Production of $K\Sigma$ final states}

\begin{figure}[b!]
\includegraphics[width=0.48\textwidth,height=0.35\textheight]{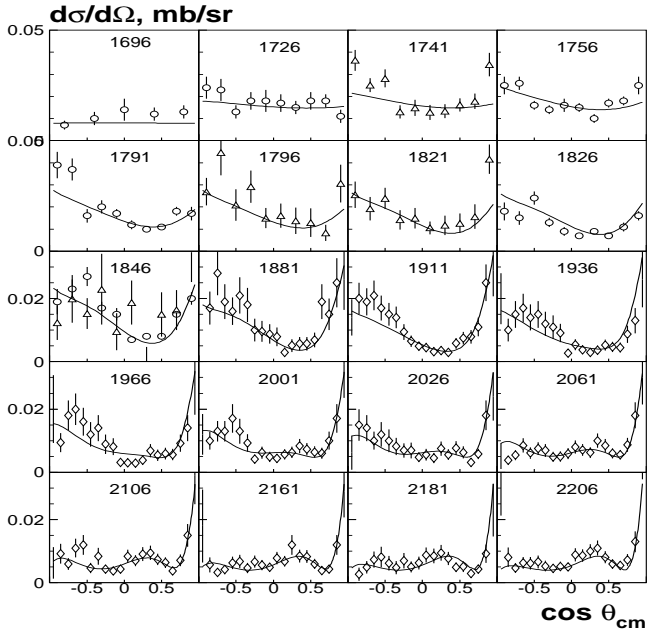}
\caption{Differential cross section of the reaction $\pi^- p\to
K^0\Sigma^0$. Curves are from the solution BnGa2011-02M. Data:
circles from Ref.~\cite{Baker:1978bb}; up triangles from
Ref.~\cite{Binford:1969ts}; diamonds from Ref.~\cite{Hart:1979jx}.}
\label{oldfig:kzszdiff1}
\end{figure}

In Figs.~\ref{oldfig:kzszdiff1} we demonstrate the quality of the
fits to the reaction $\pi^- p\to K^0\Sigma^0$ obtained with the
solution BnGa2011-02M. The data on $\pi^- p\to K^+\Sigma^-$ can be
included in the fit rather easily: only a very small adjustment of
the parameters is needed to describe them with a good quality
(solution BnGa2011-02M). As mentioned in the introduction, the high
precision photoproduction data from MAMI-C \cite{Jude:2013jzs} are
included in this analysis. Table~\ref{chis} documents the quality
for the description of the reactions with $K\Sigma$ final states.

Although the fit describes the $\pi^-p\to K^0\Sigma^0$ differential
cross section in the energy region above 1825\,MeV with a very good
$\chi^2=1.02$, it misses the structure at backward angles for the
invariant masses between 1911 and 2061 MeV. Increasing weight of the
data set in the overall fit did not resolve this problem.

The $K\Sigma$ amplitudes can have isospin $1/2$ and $3/2$. The
isospin $3/2$ amplitudes are well fixed by the $\pi^+p\to
K^+\Sigma^+$ data. An estimate of isospin $3/2$ contribution to the
total cross section for $\pi^-p\to K^0\Sigma^0$ and $\pi^-p\to
K^+\Sigma^-$ suggest that both isospin amplitudes contribute almost
equally to these reaction. $\Delta$ resonances are found to dominate
the reaction $\gamma p\to K^+\Sigma^0$ even though both isospin
states could contribute to the reaction. Based on the observed cross
section we find that the $\gamma p\to K^0\Sigma^-$ cross section
should be dominated by nucleon partial waves.
\begin{table}[t!]
\caption{\label{chis}Fit quality for fits with and without inclusion
of data on the reaction $\pi^- p\to K^+\Sigma^-$.\vspace{-2mm}}
\begin{center}
\renewcommand{\arraystretch}{1.17}
\begin{tabular}{lcccll}
\hline\hline
 Obs. &BnGa&BnGa& $N_{data}$ &Ref.&\\
      &2011-02M&2013-02&&\\
\hline
\multicolumn{6}{l}{$\pi^- p\to K^0\Sigma^0$}\\
$d\sigma/d\Omega$  &1.02 &0.69 & 220&\cite{Hart:1979jx} &(RAL) \\
$P$                &1.53 &1.21 & 85 &\cite{Hart:1979jx} &(RAL)  \\
$d\sigma/d\Omega$  &2.22 &1.91 & 95 &\cite{Baker:1978bb,Binford:1969ts} &(RAL)   \\
\hline
\multicolumn{5}{l}{$\pi^+ p\to K^+\Sigma^+$}\\
$d\sigma/d\Omega$  &1.46 &1.35 & 743&
\cite{Candlin:1982yv,Winik:1977mm,Carayannopoulos:1965,Crawford:1962zz,Baltay}&(var.) \\
$P$                &1.42 &1.48 & 351 &
\cite{Candlin:1982yv,Winik:1977mm,Carayannopoulos:1965,Crawford:1962zz,Baltay}&(var.) \\
$\beta$             &2.09 &1.89 & 7 &\cite{Candlin:1988pn} &(RAL) \\
\hline
\multicolumn{5}{l}{$\pi^- p\to K^+\Sigma^-$}\\
$d\sigma/d\Omega$  &2.45 &2.42  & 130 &
~\cite{Good:1969rb,Doyle:1968zz,Dahl:1969ap,Goussu:1966ps}&(var.)\\
\hline \multicolumn{5}{l}{$\gamma p\to K^+\Sigma^0$}\\
$d\sigma/d\Omega$ &1.30 &1.49  &1590 &\cite{Dey:2010hh} &(CLAS)\\
$d\sigma/d\Omega$ &1.45 &1.40  &1145 &\cite{Jude:2013jzs} &(MAMI)\\
$P$               &2.43 &2.17  &344  &\cite{Dey:2010hh} &(CLAS)\\
$\Sigma$          &2.45 &1.99  &42   &\cite{Lleres:2007tx} &(GRAAL) \\
$C_x$             &2.13 &2.56  &94   &\cite{Bradford:2006ba} &(CLAS)\\
$C_z$             &2.13 &2.06  &94   &\cite{Bradford:2006ba} &(CLAS)\\
\hline
\multicolumn{5}{l}{$\gamma p\to K^0\Sigma^+$}\\
 $d\sigma/d\Omega$ &3.25 &4.00 &48 & \cite{Carnahan:2003mk}  &(CLAS)\\
 $d\sigma/d\Omega$ &1.28 &1.45 &160& \cite{Lawall:2005np} &(SAPHIR)\\
 $d\sigma/d\Omega$ &0.87 &0.94 &72 & \cite{Castelijns:2007qt} &(CBT)\\
 $P$               &0.96 &0.82 &72 & \cite{Castelijns:2007qt} &(CBT)\\
 $d\sigma/d\Omega$ &0.61 &0.72 &72 & \cite{Schmieden:2012zz} &(CBT) \\
 $P$               &1.66 &1.35 &24 & \cite{Schmieden:2012zz}  &(CBT)\\
 $\Sigma$          &2.04 &1.68 &15 & \cite{Schmieden:2012zz}  &(CBT)\\
\hline\hline
\end{tabular}
\end{center}
\end{table}

\begin{figure}[pb]
\includegraphics[width=0.48\textwidth,height=0.35\textheight]{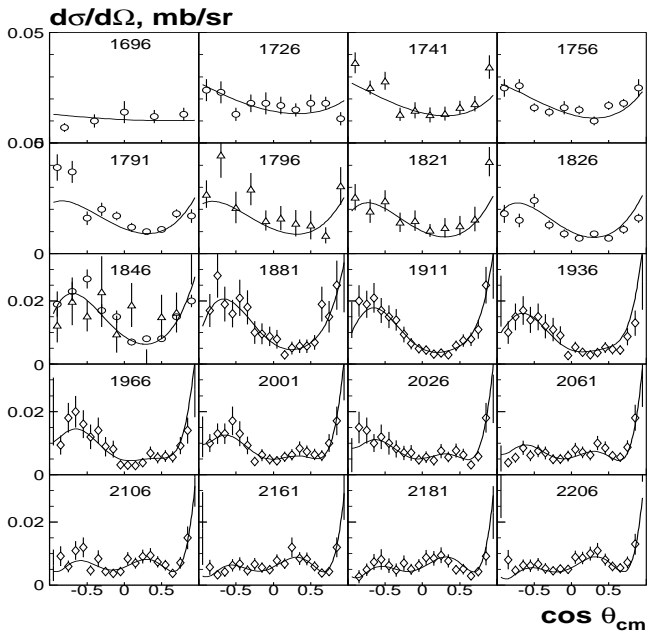}%
\caption{\label{fig:kzszdiff}Differential cross section of the
reaction $\pi^- p\to K^0\Sigma^0$.  Data: circles from
Ref.~\cite{Baker:1978bb}; up triangles from
Ref.~\cite{Binford:1969ts}; diamonds from Ref.~\cite{Hart:1979jx}.
Curves represent solution BnGa2013-02.}
\end{figure}
\clearpage
\begin{figure}[h!]
\includegraphics[width=0.48\textwidth,height=0.42\textheight]{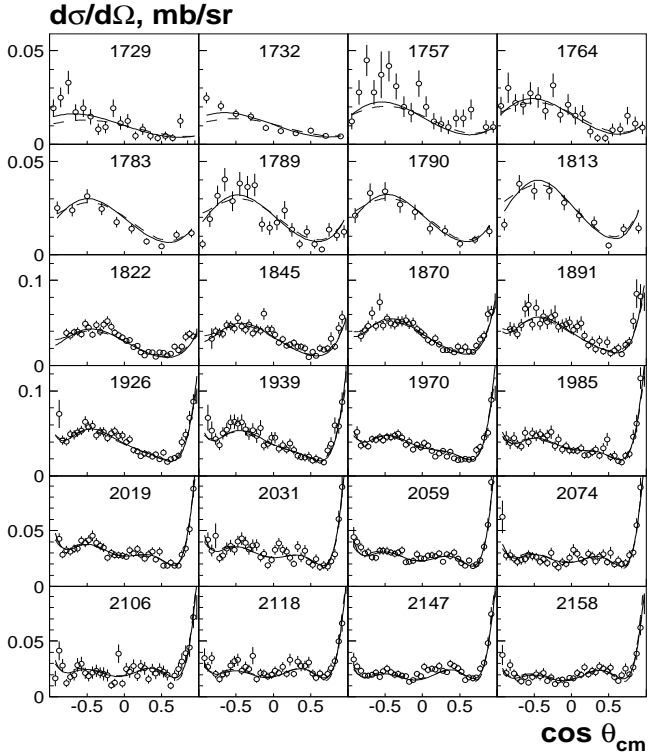}
\caption{Differential cross section for $\pi^+ p\to K^+\Sigma^+$.
The data are from~
\cite{Candlin:1982yv,Winik:1977mm,Carayannopoulos:1965,Crawford:1962zz,%
Baltay}. Full curves: the solution BnGa2013-02 and dashed curves:
BnGa2011-02M.}
\label{fig:kpspdiff}
\end{figure}
It can hence be expected that the $\pi^-p\to K^0\Sigma^0$ reaction
is sensitive to the interference between isospin $3/2$ and $1/2$
amplitudes. To check this interference we have changed the sign of
the couplings of all  nucleon resonances to the $K\Sigma$ channel.
Let us note that the description of reactions in which nucleon or
$\Delta$ resonances are dominant undergo little changes only. In
those reactions, the interference is small and the $K\Sigma$
amplitudes contribute mostly quadratically.

A fit with all ($N\to K\Sigma$) K-matrix coupling constants reversed
cured the problems in the description of the angular distributions
in Fig.~\ref{oldfig:kzszdiff1}. The solution also describes
acceptably well all other reactions with $K\Sigma$ final states but
introduces small changes of the properties like masses and widths of
baryon resonances. As a result, the overall description of the data
became worse.

These findings initiated a full systematic study of $N\to K\Sigma$
decay amplitudes changing the signs of all K-matrix coupling
(resonances and background terms) in all possible combinations. The
relative signs of coupling constants within a given partial wave
turned out to be well defined; mostly, the signs of the full partial
wave amplitudes needed to be changed. The optimum was found when the
sign was changed for the $S_{11}$, $D_{13}$ and $F_{15}$ partial
waves. This fit produced an overall likelihood value which was about
740 better than the one in solution BnGa2011-02M. We will denote
this solution as BnGa2013-02. It describes the high energy
$\pi^-p\to K^0\Sigma^0$ data with the $\chi^2=0.69$. The description
of the differential cross section with this solution is shown in
Fig~\ref{fig:kzszdiff}.
\begin{figure}[h!]
\includegraphics[width=0.48\textwidth,height=0.42\textheight]{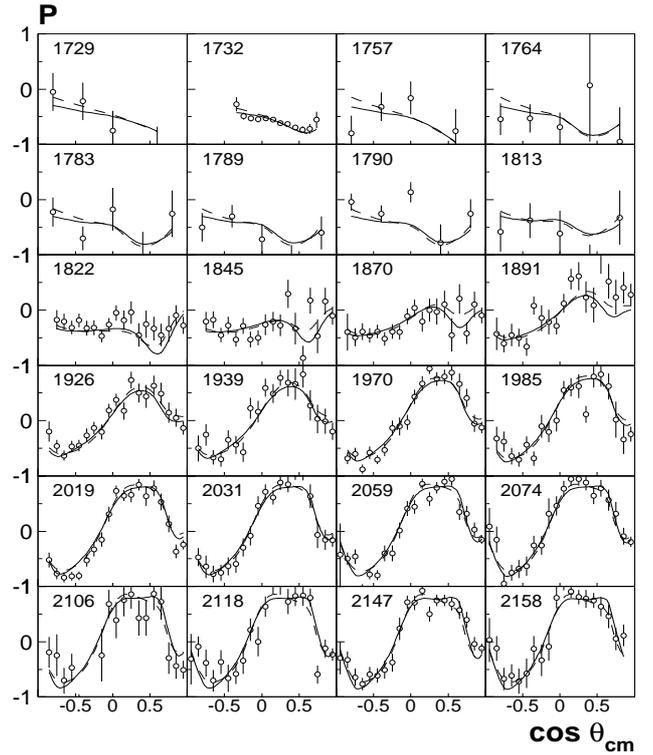}
\caption{Recoil asymmetry from $\pi^+ p\to K^+\Sigma^+$. The data
are from~
\cite{Candlin:1982yv,Winik:1977mm,Carayannopoulos:1965,Crawford:1962zz,%
Baltay,Bellamy:1972fa}. Full curves: the solution BnGa2013-02 and
dashed curves: BnGa2011-02M.\vspace{5mm}}
\label{fig:kpsppola1}
\end{figure}
\begin{figure}[ht]
\includegraphics[width=0.48\textwidth,height=0.17\textheight]{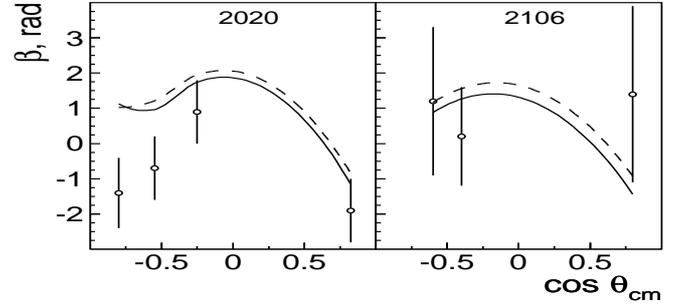}%,height=0.15\textheight
\caption{Spin-rotation parameter $\beta$ from the reaction $\pi^+
p\to K^+\Sigma^+$. The data are from~\cite{Candlin:1988pn}. Full
curves: the solution BnGa2013-02 and dashed curves: BnGa2011-02M.
Note that $\beta$ is $2\pi$ cyclic.}
 \label{fig:kpspbeta}
\end{figure}

\begin{figure}[ht]
\includegraphics[width=0.48\textwidth,height=0.21\textheight]{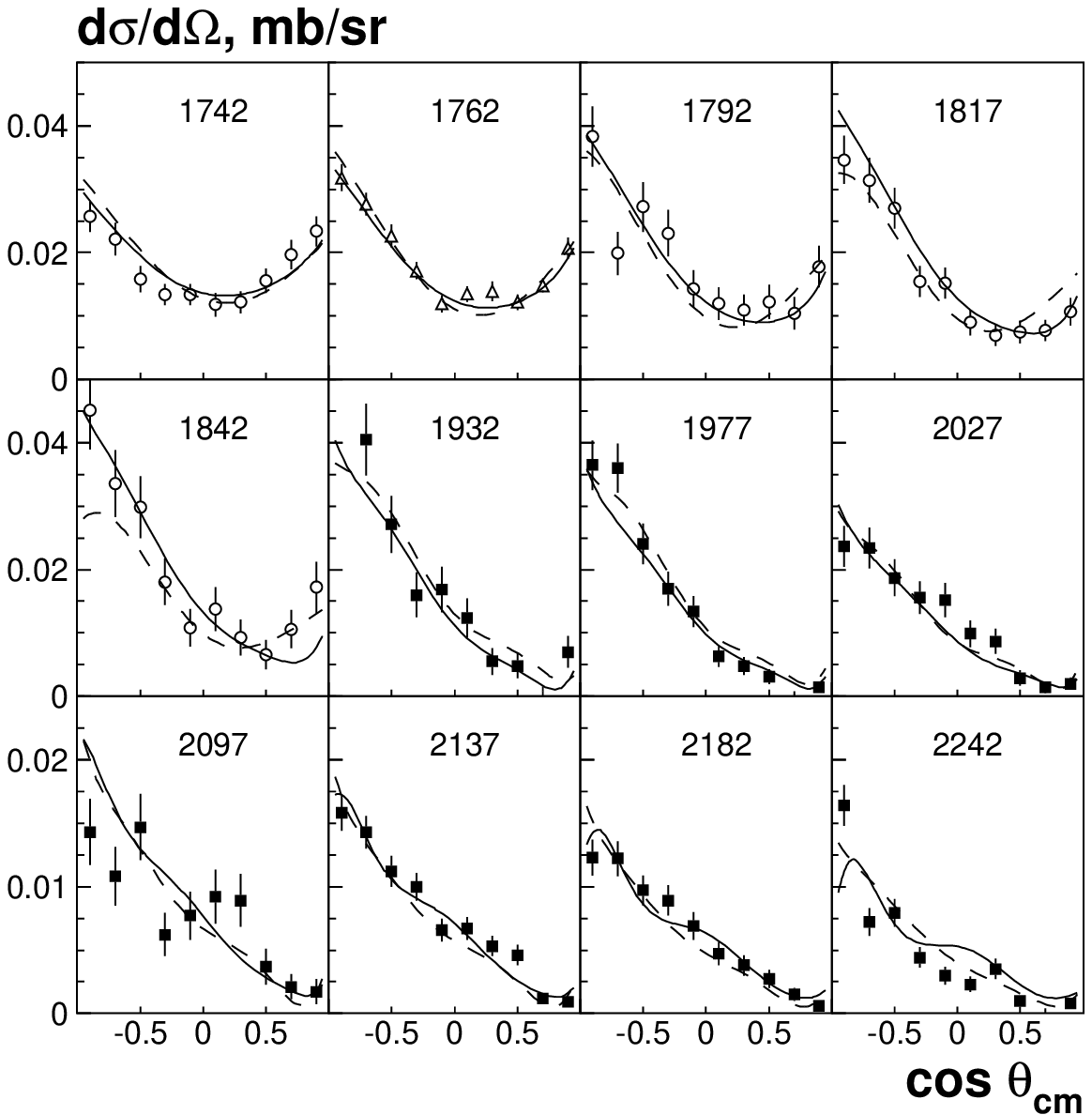}
\caption{Differential cross section of the reaction $\pi^- p\to
K^+\Sigma^-$. Full curves: the solution BnGa2013-02 and dashed
curves: BnGa2011-02M. The data are from ~\cite{Good:1969rb}
(circles); \cite{Doyle:1968zz} (up triangles ); \cite{Dahl:1969ap}
(squares ); \cite{Goussu:1966ps} (diamonds).\vspace{3mm}}
\label{fig:kpsmdiff1}
%\end{figure}
%\begin{figure}[ht]
\includegraphics[width=0.48\textwidth,height=0.35\textheight]{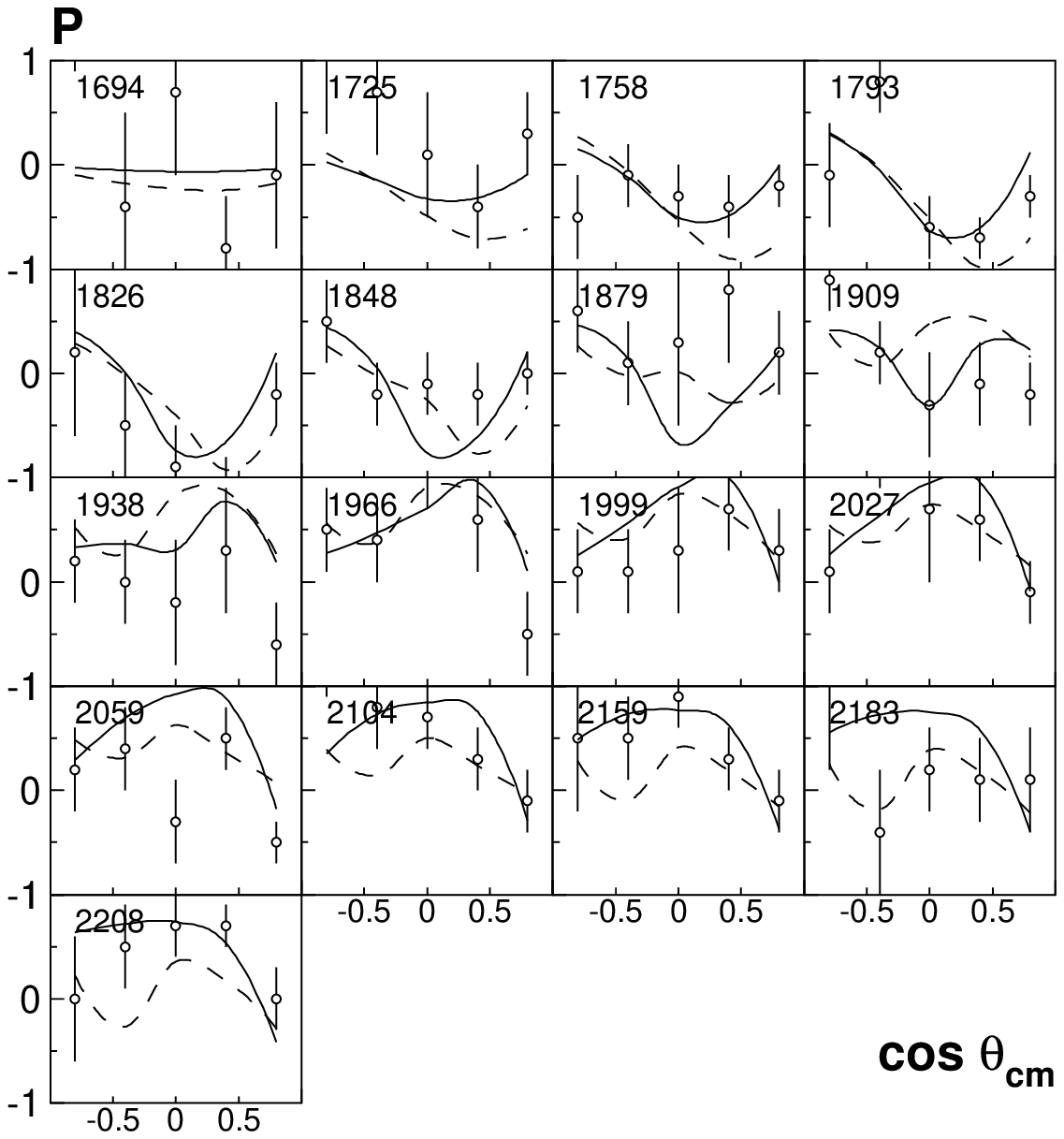}
\caption{Recoil asymmetry in the reaction $\pi^- p\to K^0\Sigma^0$.
Full curves: the solution BnGa2013-02 and dashed curves:
BnGa2011-02M. The data are from~\cite{Hart:1979jx}.}
%; diamonds from Ref.~\cite{Hart:1979jx}
\label{fig:kzszpola}
\end{figure}

The improvement in the description of the data with $K\Sigma$ final
states is not very impressive but noticeable as can be seen in
Table~\ref{chis} and in a few figures. These show for $\pi^+p\to
K^+\Sigma^+$ the differential cross section
(Fig.~\ref{fig:kpspdiff}), the recoil asymmetry
(Fig.~\ref{fig:kpsppola1}), and the spin rotation parameter
(Fig.~\ref{fig:kpspbeta}).  For $\pi^-p\to K^+\Sigma^-$ we show the
differential cross section (Fig.~\ref{fig:kpsmdiff1}) and for
$\pi^-p\to K^0\Sigma^0$ the recoil asymmetry
(Fig.~\ref{fig:kzszpola}). In the figures, solution BnGa2011-02M is
shown with dashed lines and BnGa2013-02 with full lines.

The solution BnGa2013-02 describes the backward structure in the
$\pi^-p\to K^0\Sigma^0$ reaction much better (see
Fig.~\ref{fig:kpspdiff}) and even provides a better description of
the recoil asymmetry. However, the data are not really enforcing the
changes which were introduced. The differential cross section for
this reaction is described equally well by both solutions.
Measurements of the rotation parameter for $\pi^-p\to K^+\Sigma^-$
and/or $\pi^-p\to K^0\Sigma^0$ could be crucial to discriminate
these solutions.

\begin{figure}[pt]
\includegraphics[width=0.48\textwidth,height=0.52\textheight]{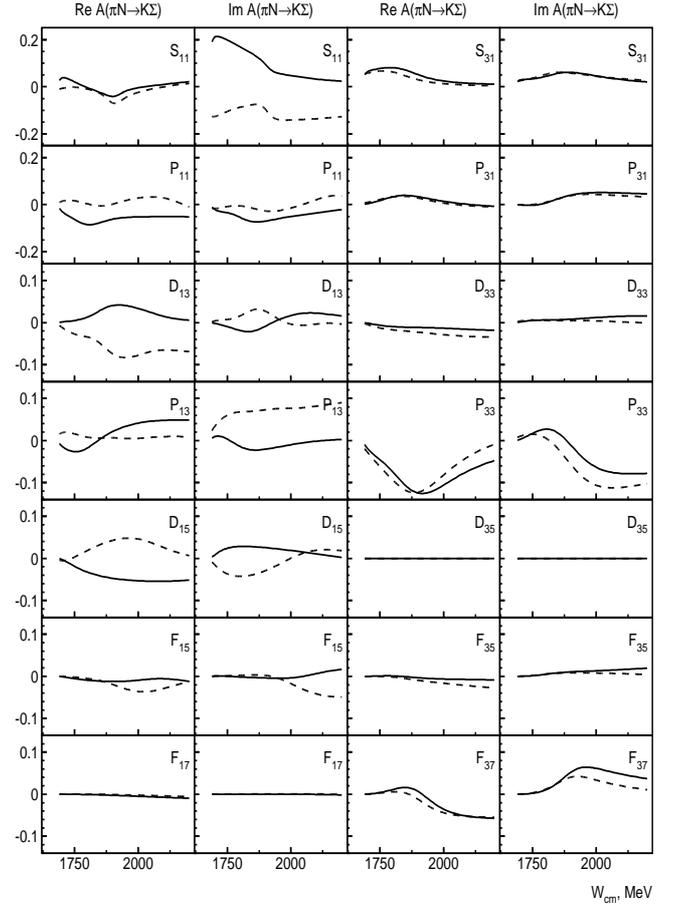}
\caption{The unitary partial $\pi N\to K\Sigma$ amplitudes. The
solution BnGa2013-02 is shown with full curves and BnGa2011-02M with
dashed curves.}
\label{fig:par_waves}
\end{figure}

The new fit BnGa2013-02 describes the $\gamma p\to K^+\Sigma^0$
differential cross section measured by the CLAS collaboration
\cite{Dey:2010hh} slightly worse. However it describes better most
polarization observables (apart from $C_x$) and the differential
cross section measured at MAMI-C \cite{Jude:2013jzs}. However, in
general the fits of these data by BnGa2011-02M or BnGa2013-02 are of
similar quality. Similar changes are observed in the description of
the reaction $\gamma p\to K^0\Sigma^+$: the solution BnGa2013-02
describes differential cross section slightly worse but polarization
observables better. With the presently available data, there is no
unambiguous decision in favor of BnGa2011-02M or BnGa2013-02. The
differences between the two solutions can be visualized by comparing
the partial wave amplitudes, see Fig.~\ref{fig:par_waves}. All
isospin $3/2$ partial waves are very close in both solutions while
nucleon partial waves differ in sign.

\section{Total cross sections and partial wave contributions}
\begin{figure}[pt]
\begin{tabular}{cc}
\includegraphics[width=0.24\textwidth]{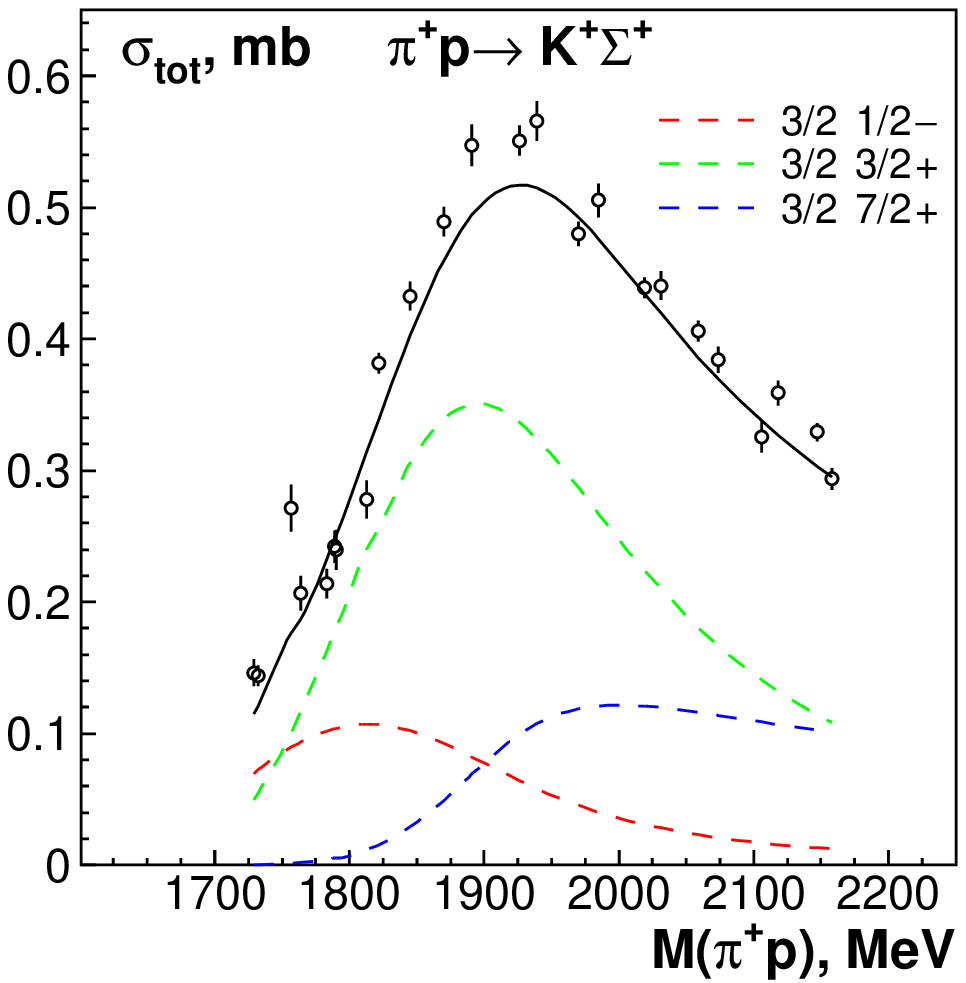}&
\hspace{-3mm}\includegraphics[width=0.24\textwidth]{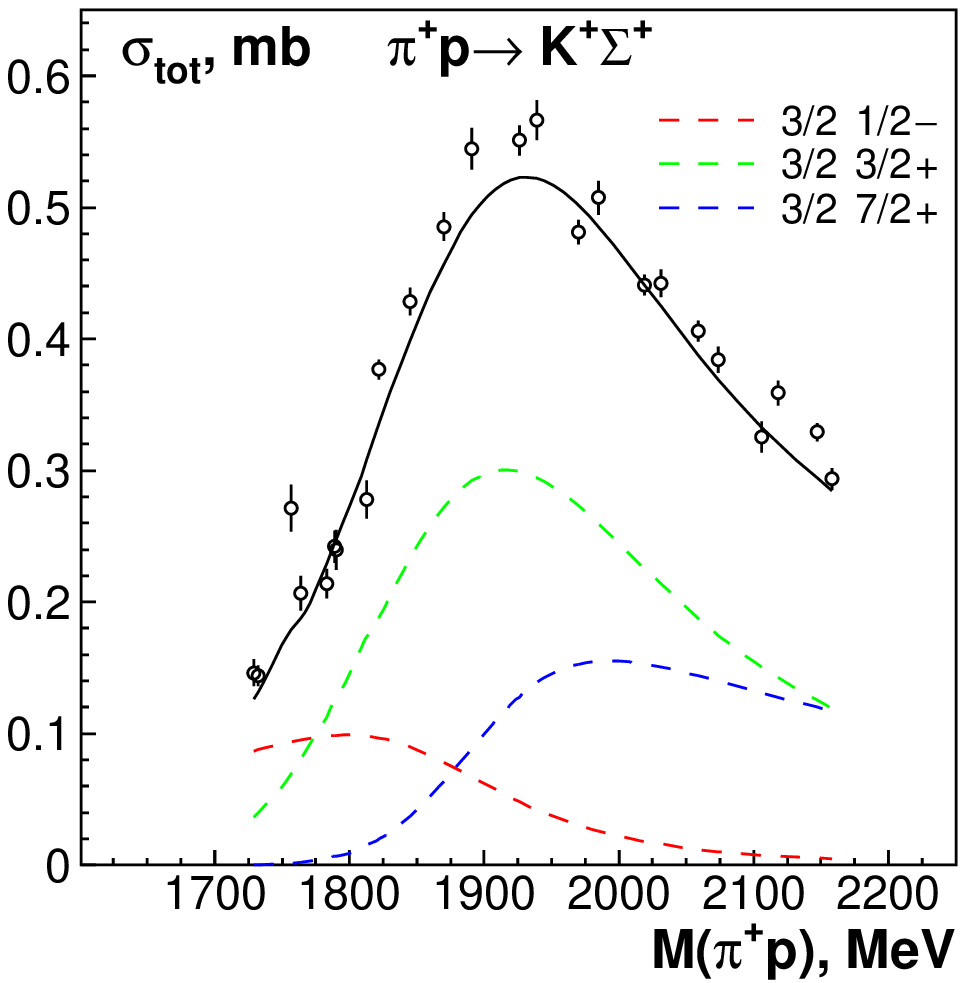}\\
\includegraphics[width=0.24\textwidth]{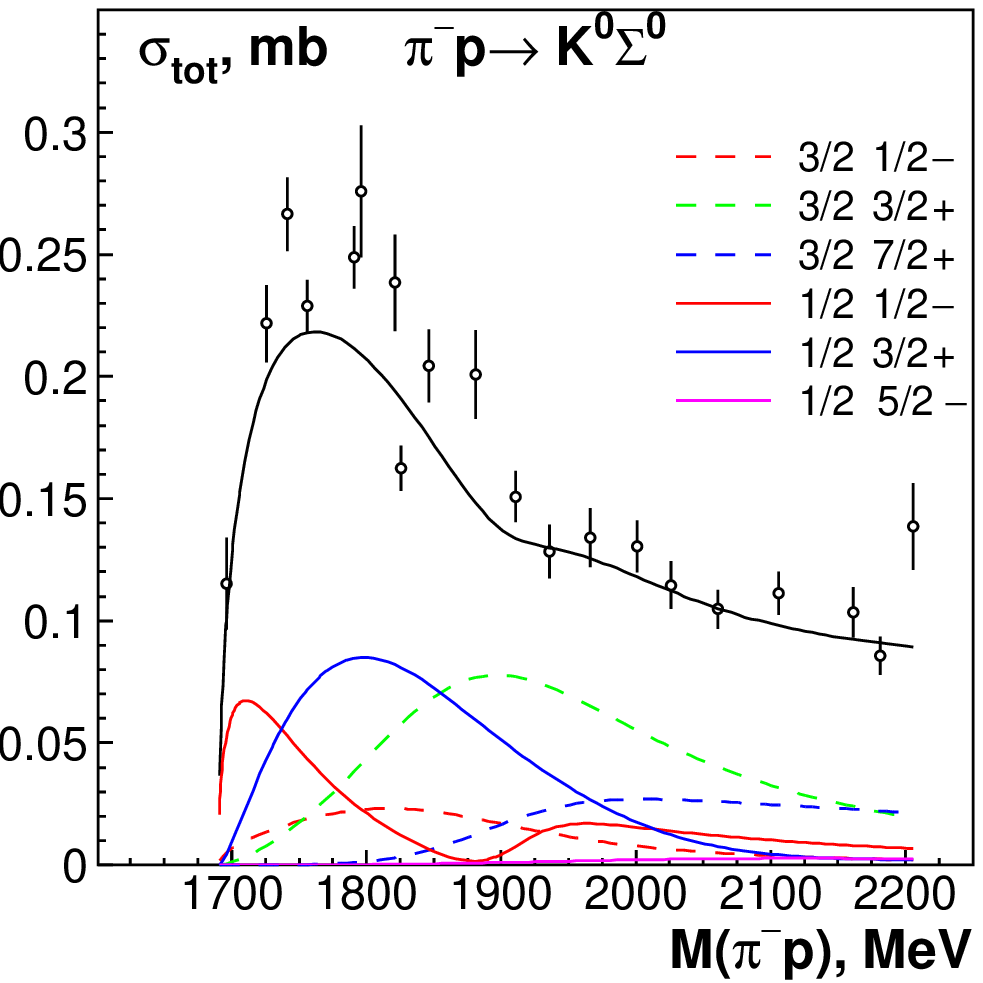}&
\hspace{-3mm}\includegraphics[width=0.24\textwidth]{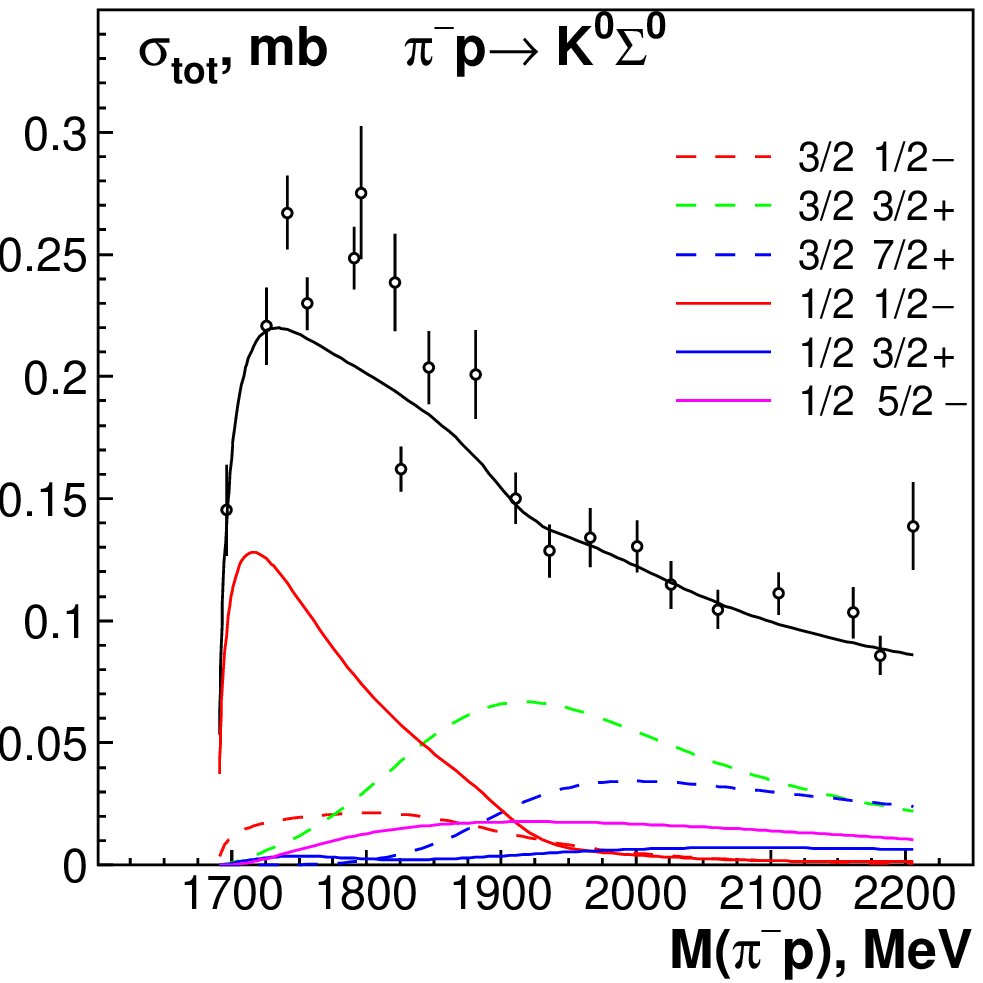}\\
\includegraphics[width=0.24\textwidth]{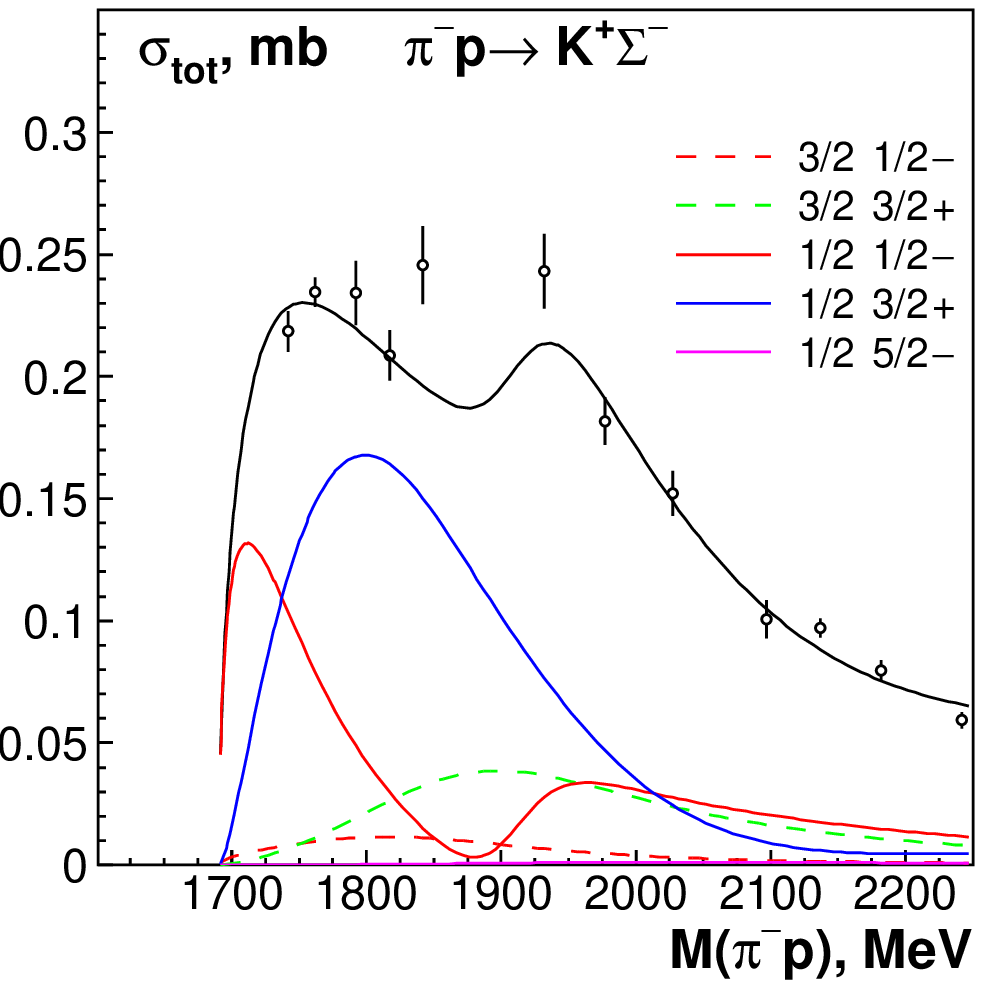}&
\hspace{-3mm}\includegraphics[width=0.24\textwidth]{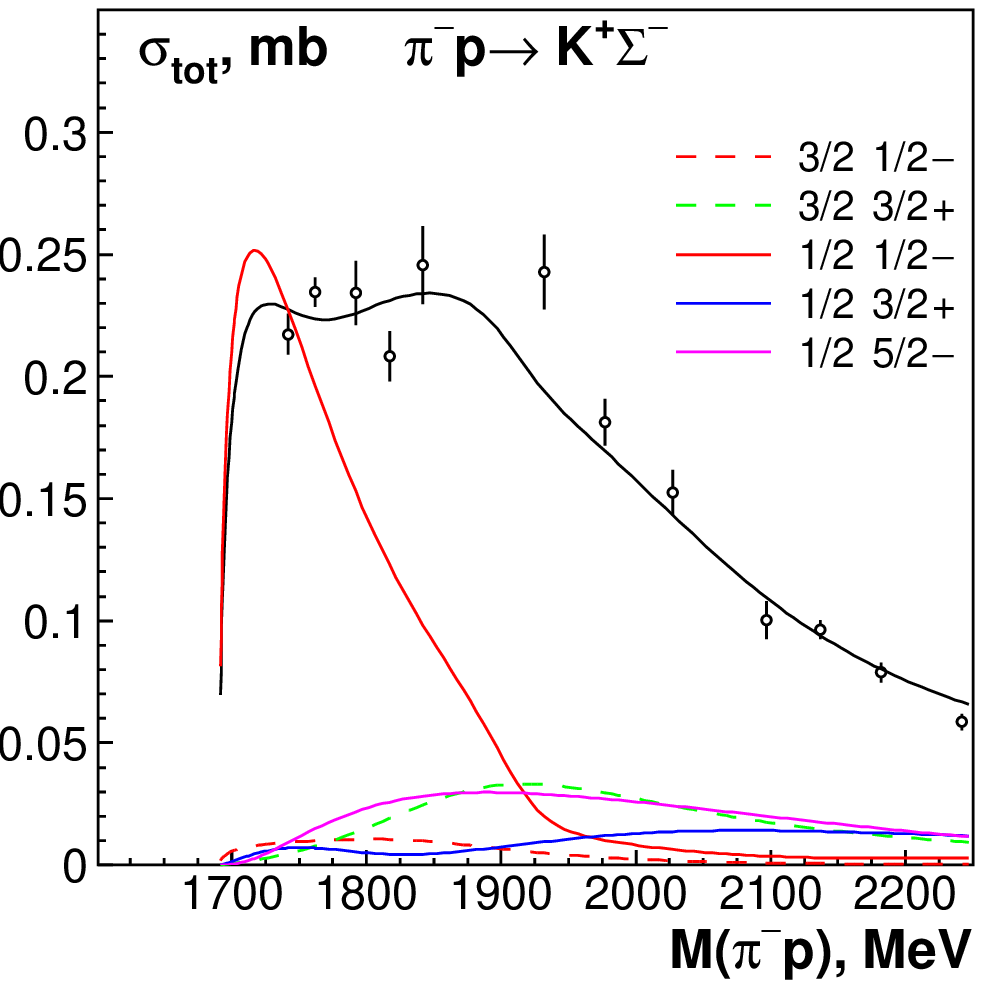}
\end{tabular}
\caption{Partial wave contributions to the total cross section of
$\pi^+ p\to K^+ \Sigma^+$, $\pi^- p\to K^0 \Sigma^0$ and $\pi^- p\to
K^+ \Sigma^-$. Data points represent the summation over the full
angular range. Left-side panels show the solution BnGa2011-02M and
right-side panels BnGa2013-02.}
\label{fig:k0s0tot}
\end{figure}

The partial wave contributions to the total cross section for the
three $\pi N\to K\Sigma$ reactions derived in fits BnGa2011-02M and
BnGa2013-02, respectively, are shown in Fig.~\ref{fig:k0s0tot}. The
contributions of isospin $3/2$ partial waves hardly changed. Both
solutions are well within the boundaries of the systematic error
defined for solution BG2011-02. However, the contributions of
nucleon resonances have undergone significant changes. In the
solution BnGa2011-02M the dominant contribution to $\pi^-p\to
K^0\Sigma^0$ and $\pi^- p\to K^+\Sigma^-$ comes from the $P_{13}$
partial wave while in the solution BnGa2013-02, this wave is very
weak. The $S_{11}$ contribution has become stronger by a factor 2 in
BnGa2013-02 and in the $\pi^-p \to K^+\Sigma^-$ it is the dominant
partial wave. Moreover, in BnGa2011-02M, destructive interference is
observed in the region of $N(1895)S_{11}$. This destructive
interference is compensated by a large intensity from the $P_{13}$
partial wave which reaches a maximum at 1800\,MeV. In BnGa2013-02,
the $S_{11}$ wave is rather smooth in the region 1900 MeV, although
a small contribution from $N(1895)S_{11}$ interfering constructively
within the $S_{11}$ wave improves the fit.

\section{Resolving the ambiguity in $K\Sigma$ amplitudes}

\begin{figure}[pt]
\includegraphics[width=0.48\textwidth]{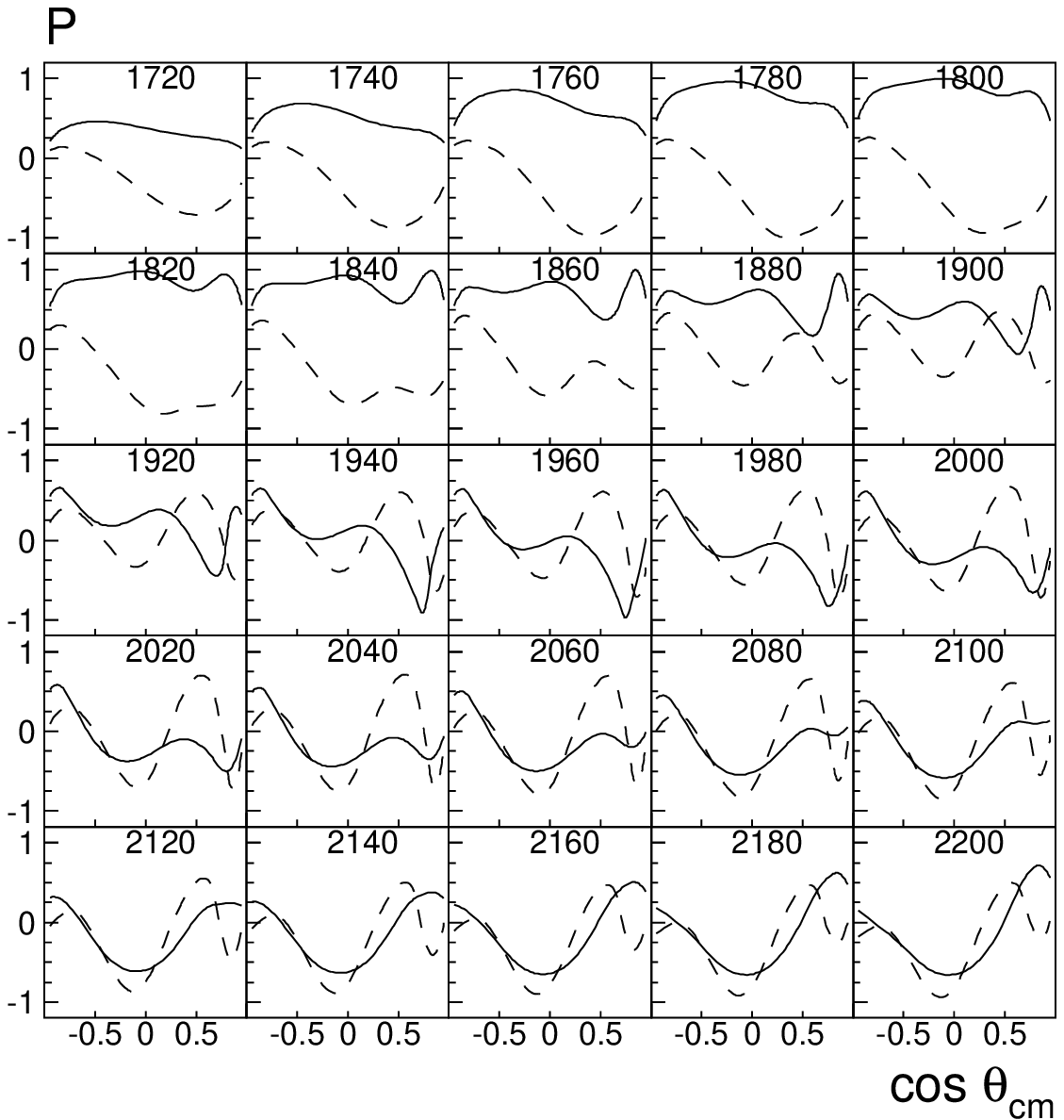}
\caption{Prediction of the recoil asymmetry in $\pi^- p\to
K^+\Sigma^-$ for solution BnGa2013-02 (full curves) and BnGa2011-02M
(dashed curves).}
\label{fig:kpsm_p}
%\end{figure}
%\begin{figure}[ph]
\includegraphics[width=0.48\textwidth]{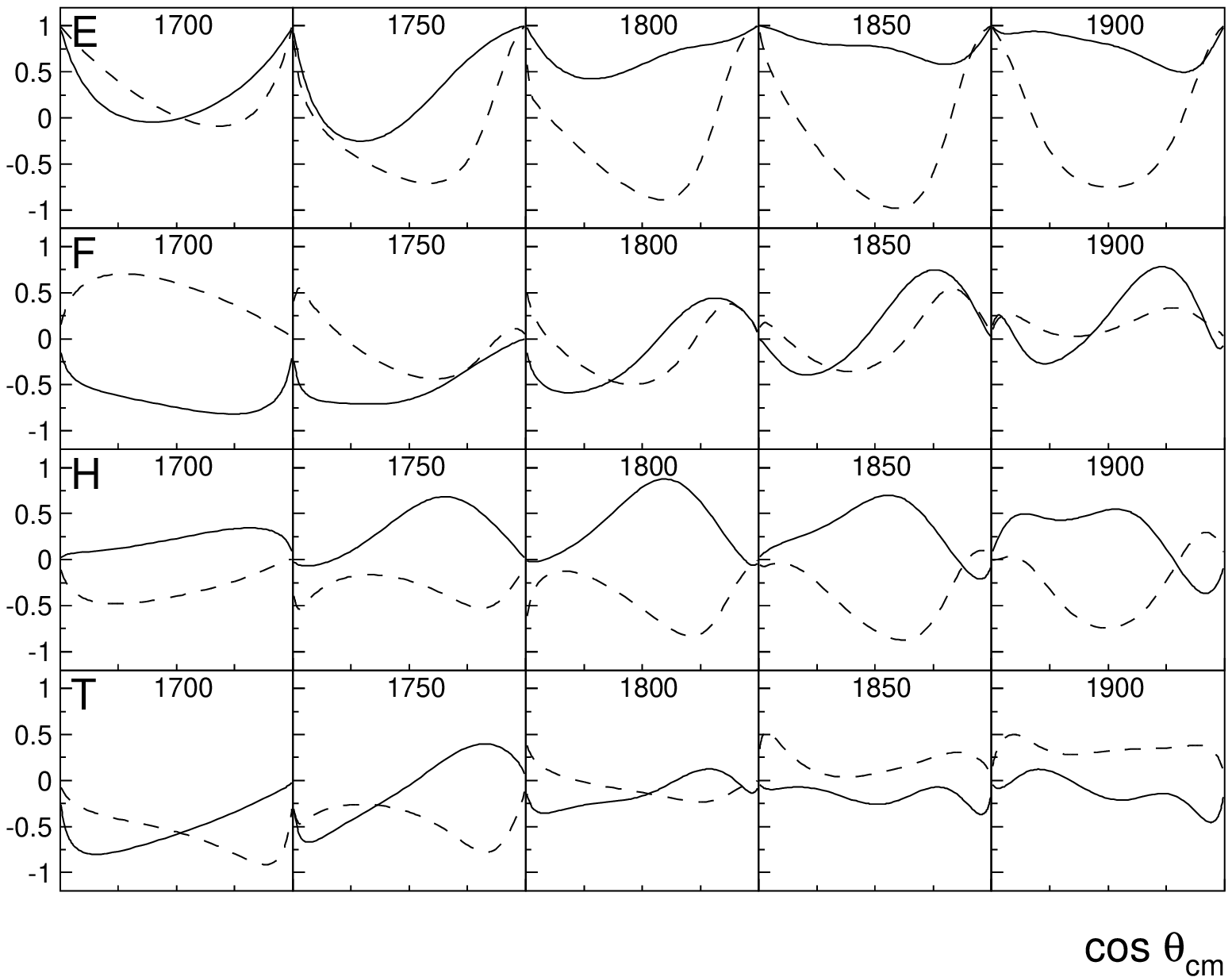}
\caption{Prediction for the polarization observables of the $\gamma
p\to K^0\Sigma^0$ reaction. The solution BnGa2013-02 is shown with
full curves and BnGa2011-02M with dashed curves.}
\label{fig:gp_pred}
\end{figure}

As mentioned above the present ambiguity is a consequence of the
lack of data on polarization observables from pion and photo-induced
reactions. In Fig.~\ref{fig:kpsm_p} we show predictions for the
recoil asymmetry for $\pi^- p\to K^+\Sigma^-$. In the 1750-1900\,MeV
region, the two solutions both predict large asymmetries but
different in sign. An additional measurement of the spin rotation
parameter would provide a full data base which would define the
contributions from all leading partial waves unambiguously. Data on
the recoil asymmetry for $\pi^- p\to K^+\Sigma^-$ could be measured
at GSI by the HADES collaboration.

The data on the $\gamma p\to K^0\Sigma^+$ is another important
source of information which can resolve this ambiguity. The
prediction for target asymmetry and double polarization observables
are shown in Fig~\ref{fig:gp_pred}. These data can be obtained by
the CBELSA/TAPS and CLAS collaborations and will not only help to
resolve this ambiguity but also to define resonances in the 2\,GeV
region more firmly.

\section{Comparison with other work}
Coupled channel analyses of pion and photo-induced production of the
$K \Sigma$ final states have been carried out by several groups.
Here we discuss only recent results.

\begin{figure*}[pt]
\bc
\begin{tabular}{ccc}
\includegraphics[height=0.24\textwidth,width=0.32\textwidth]{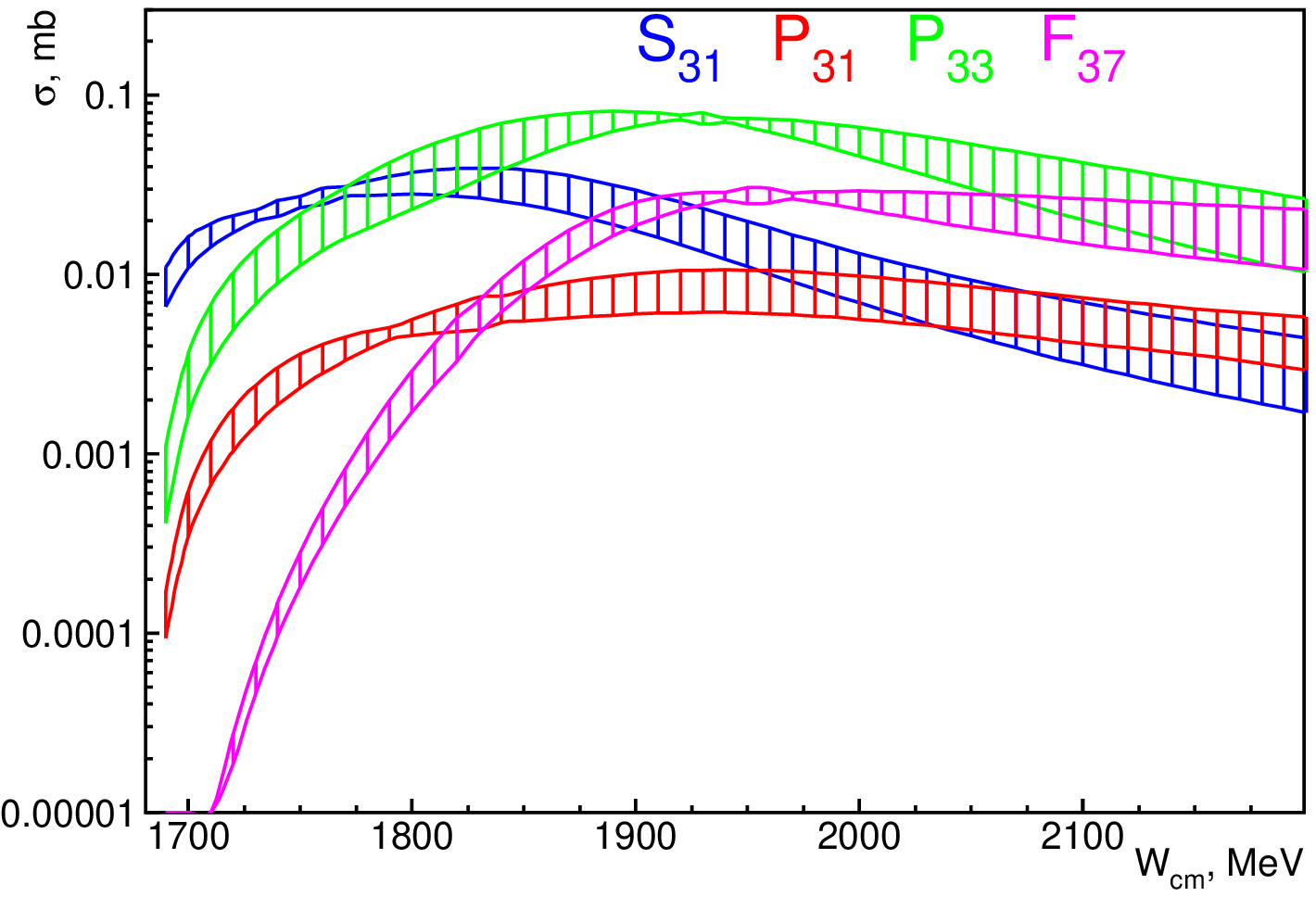}&
\hspace{-4mm}\includegraphics[height=0.24\textwidth,width=0.32\textwidth]{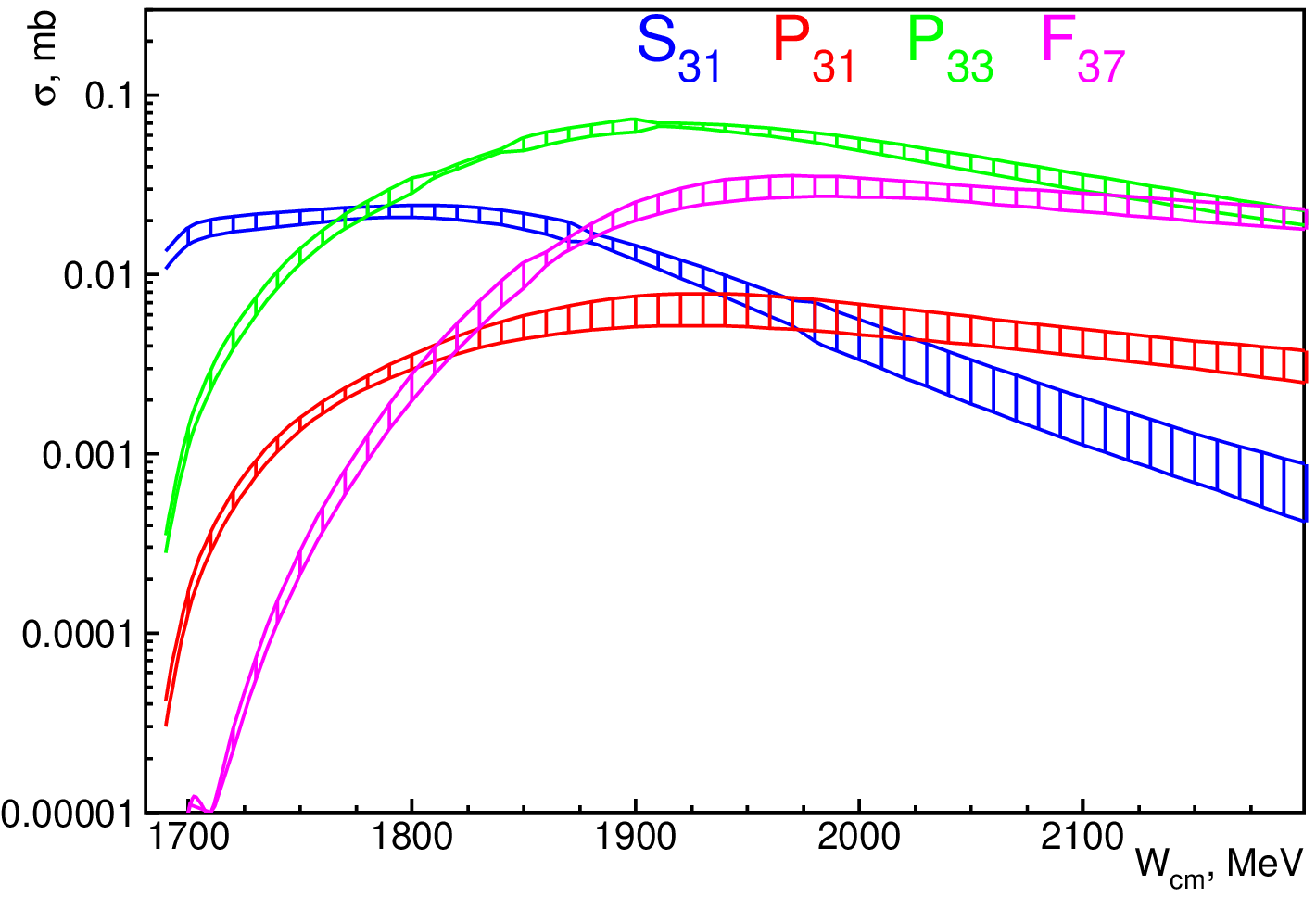}&
\hspace{-4mm}\includegraphics[height=0.237\textwidth]{total_cs_KzSz_pcs_I=3h.eps}\\
\includegraphics[height=0.24\textwidth,width=0.32\textwidth]{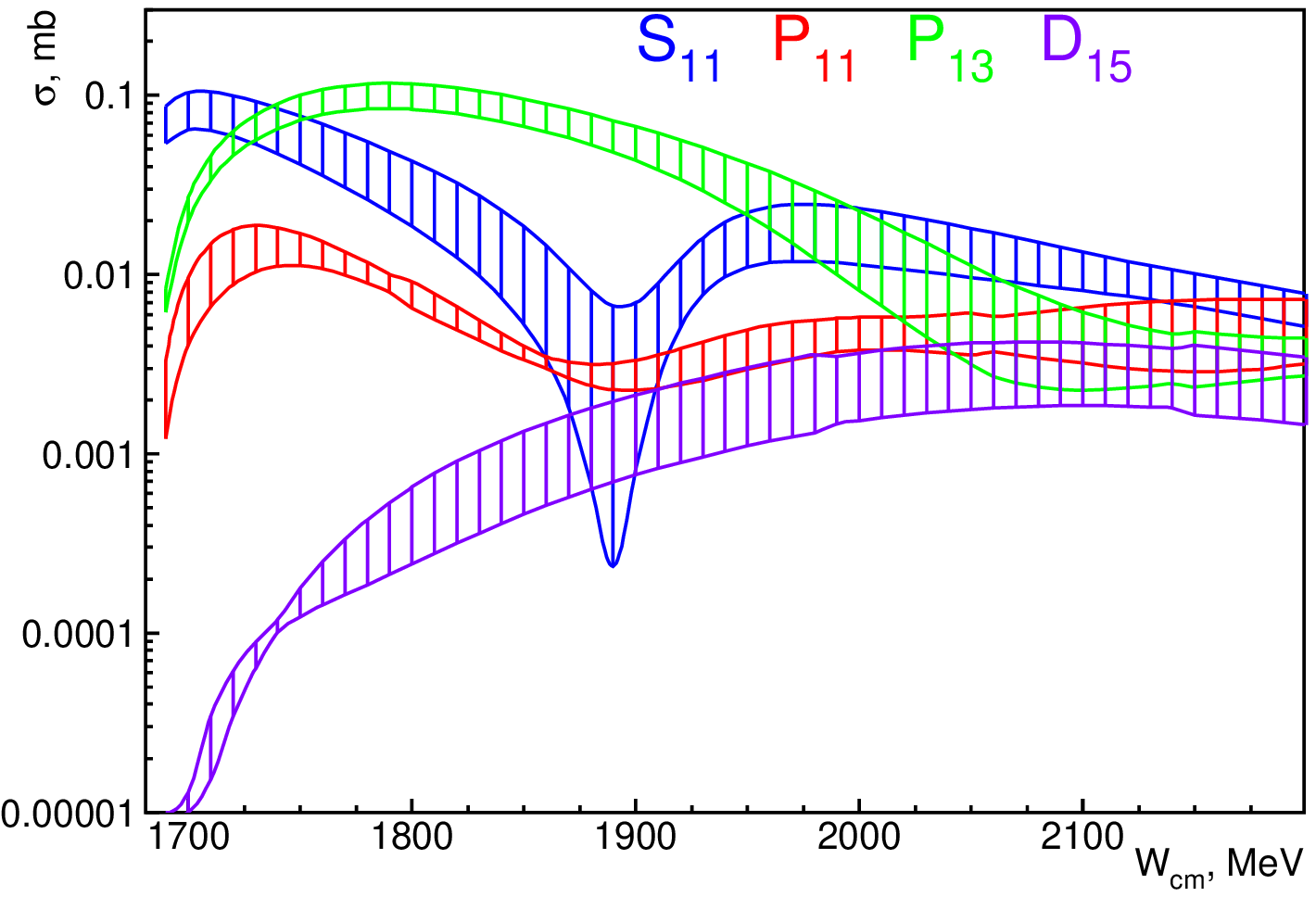}&
\hspace{-4mm}\includegraphics[height=0.24\textwidth,width=0.32\textwidth]{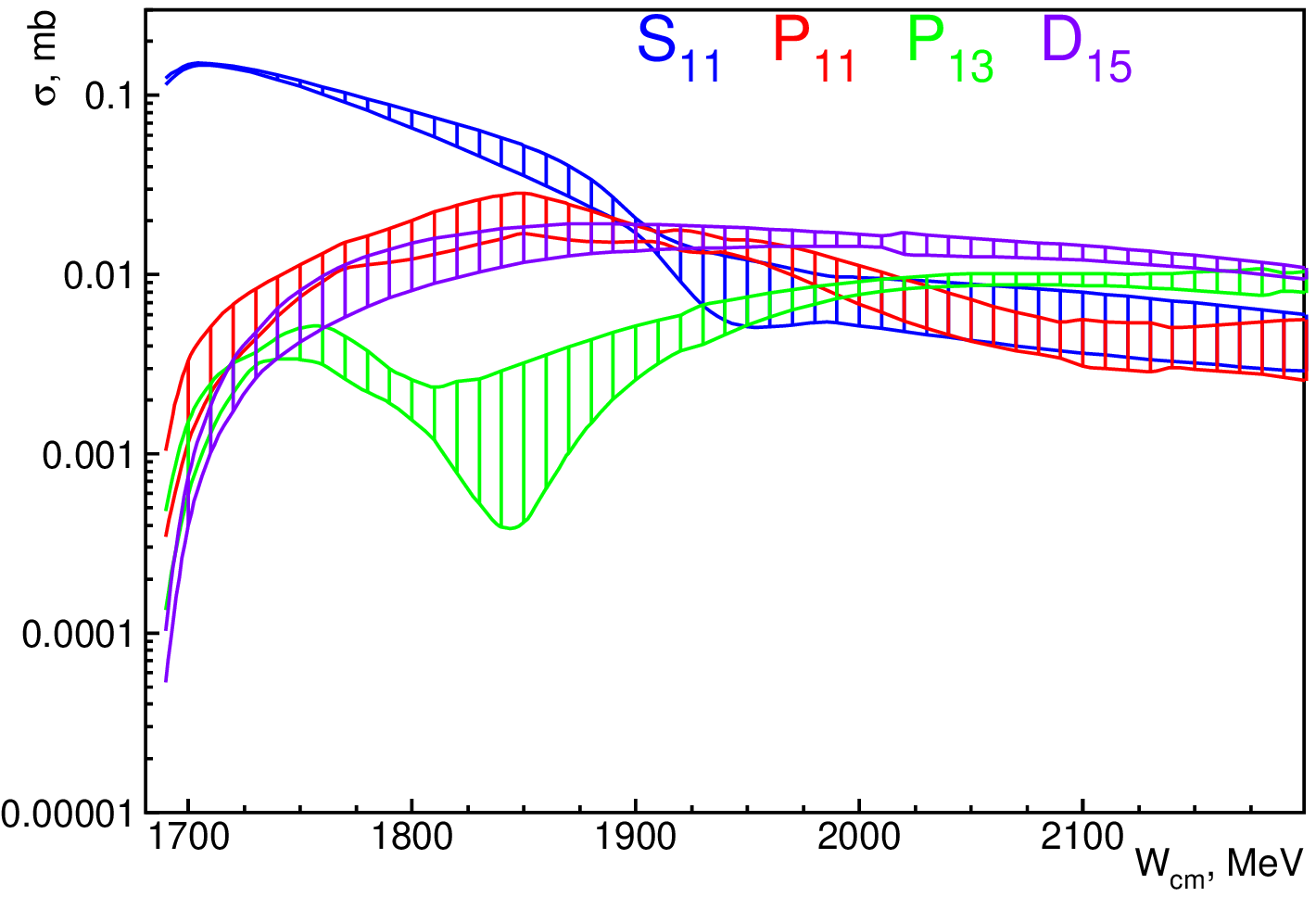}&
\hspace{-4mm}\includegraphics[height=0.237\textwidth]{total_cs_KzSz_pcs_I=1h.eps}
\end{tabular}
\ec
\caption{\label{fig:contributions}Partial cross section
contributions for the reaction $\pi^- p\to K^0 \Sigma^+$. Left-side
panels: Solution BnGa2011-02M, center: solution BnGa2013-02,
right-side panels: Bonn-J\"ulich solution \cite{Ronchen:2012eg}. In
the Bonn-Gatchina solutions only the most significant contributions
are shown.}
\end{figure*}

The Osaka-Tokyo-Argonne group \cite{Kamano:2013iva} and the Gie\ss
en group \cite{Cao:2013psa} do not report partial wave contributions
to the cross sections, hence we restrict the discussion to the
results from Bonn-J\"ulich \cite{Doring:2010ap,Ronchen:2012eg} and
Bonn-Gatchina
\cite{Anisovich:2011ye,Anisovich:2011fc,Anisovich:2012ct}.

The first Bonn-J\"ulich fit (Bonn-J\"ulich 2010)
\cite{Doring:2010ap} to the $\pi^+ p\to K^+\Sigma^+$ data and the
Bonn-Gatchina fit BnGa2011-02 \cite{Anisovich:2011ye} for the
isospin $3/2$ sector gave significantly different answers concerning
the magnitude of the most significant partial wave contributions. In
the Bonn-J\"ulich analysis~\cite{Doring:2010ap}, the largest
contribution to the cross section is assigned to the $(I,J^P)=(3/2,
1/2^-)$ wave, followed by a much smaller $(3/2, 3/2^+)$
contribution. The $(3/2, 7/2^+)$ contribution starts at threshold,
exceeds the $(3/2, 3/2^+)$ contribution above 2\,GeV but stays
always well below the $(3/2, 1/2^-)$ wave. In the Bonn-Gatchina fit
BnGa2011-02M (as well as in BnGa2013-02), the $(3/2, 3/2^+)$
contribution is by far dominant at 1900\,MeV and falls off at higher
energies. The $(3/2, 1/2^-)$ wave provides a significant but much
smaller contribution. The $(3/2, 7/2^+)$ contribution rises slowly
with energy, adopts the same height as $(3/2, 1/2^-)$ contribution
at 1.9\,GeV, and becomes the largest contribution at the highest
energy.

In the most recent analysis of the Bonn-J\"ulich group
\cite{Ronchen:2012eg} a new solution was found. This solution uses a
much larger data base and includes pion-induced reactions with
different $K\Sigma$ final states. For sake of convenience, we show
in Fig.~\ref{fig:contributions} the results of three analyses,
BnGa2011-02M and BnGa2013-02, and of Bonn-J\"ulich 2012. In the
isospin $3/2$ sector, the three analyses identify the same partial
waves, $P_{33}$, $S_{31}$, and $F_{37}$, as dominant contributions,
even though in both Bonn-Gatchina analyses the $S_{31}$ falls off
with energy while it very slowly rises in the Bonn-J\"ulich
analysis. Only in the smaller contributions significant differences
can be found: in particular the Bonn-Gatchina analysis does not find
a significant contribution from the $D_{35}$ wave, and also possible
contributions from the $G_{37}$ and $G_{39}$ waves are fitted to
zero. Instead, more intensity is assigned to the leading $P_{33}$
wave.

The contributions in $I=1/2$ sector are less well defined. Since
both $N$ and $\Delta$ resonances contribute with similar strengths
to the reaction $\pi^-p\to K^0\Sigma^0$, and uncertainties on the
sign of $K\Sigma$ coupling constants of $N$ resonances play a large
role. Thus, even the leading partial waves are different in
BnGa2011-02M and BnGa2013-02. However, there is fair or even good
agreement between the leading partial waves of BnGa2013-2 and
Bonn-J\"ulich 2012: the $S_{11}$ wave is leading at low energies and
$D_{15}$ and $P_{13}$ become important at the highest energy.
$P_{11}$ is important in both analyses even though more pronounced
in BnGa2013-02. Smaller contributions are present in both analyses
even though their strengths may differ: BnGa assigns more intensity
to the $F_{15}$ wave, Bonn-J\"ulich to $F_{17}$. There are
significant $D_{13}$ contributions already at low energy in
BnGa2013-02, a partial wave which gives a contribution that rises
slowly with energy in the Bonn-J\"ulich analysis.

In general, the Bonn-J\"ulich solution varies more smoothly as a
function of energy, the Bonn-Gatchina solution has more structure.
The reason for this difference is due to the larger number of
resonances used in the Bonn-Gatchina analysis which fits not only
pion-induced reactions but also photo-induced reactions. They are of
considerably higher statistical power and require introduction of
more resonances.

In the comparison, one has to have in mind that the ``area of
uncertainty" or error bands are derived differently in the two
approaches. The Bonn-J\"ulich group has used two different model
assumptions yielding two sets of contributions to the cross section
(solid and dashed curves). The Bonn-Gatchina group has used in total
eleven different parameterizations of partial waves and/or different
weight factors (for BnGa2011-02M and BnGa2013) which all gave
acceptable fits to the data. Errors are defined from the variance of
the respective contributions.

\section{Conclusion}

The investigation of the reactions with $K\Sigma$ final state
revealed a discrete ambiguity in the sign of the $K\Sigma$ coupling
constants of leading nucleon partial wave amplitudes. While the
isospin $3/2$ amplitudes are firmly defined by the $\pi^+p \to
K^+\Sigma^+$ and $\gamma p\to K^+\Sigma^0$ data (where isospin $3/2$
contributions play a dominant role), the lack of polarization data
in other reactions with $K\Sigma$ final states leads to two rather
different solution. Both solutions provide a  good overall
description of the present data base. The solution BnGa2011-02M
(which is very similar to the BaGa2011-02 solution) misses some
structures in $\pi^-p\to K^0\Sigma^0$ data; the new solution
BnGa2013-02 describes those data better but the overall fit is not
really superior. Even though we give preference to the new solution,
the decision which one of the two solution is closer to the truth
can only be made when new data on polarization observables are
available, either from reaction $\pi^-p\to K^+\Sigma^-$ or from
$\gamma p\to K^0\Sigma^0$. For both reactions predictions were made
how to discriminate the two solutions. It is shown that a
measurement of the recoil polarization for the former reaction or a
measurement of a polarization variable like $E$, $F$, $H$, or $T$
for the latter reaction will be sufficient to resolve the ambiguity.

The new solution compares favorably to the solution obtained by the
Bonn-J\"ulich group. In the isospin $1/2$ sector, the comparison
shows even striking similarity of the leading waves. In the isospin
$3/2$ sector, the leading waves are similar even though minor waves
differ in the magnitude of their contributions to the respective
cross sections. The similarity of the results carries an important
message: the Bonn-Gatchina and Bonn-J\"ulich groups use rather
different analysis methods, but the leading waves are very similar.
Obviously, the leading waves are defined by the data base, but not
by the method. Differences in details are, however, important and
ask for continued efforts, both in augmenting the data base and in
the development of analysis techniques.

\section*{Acknowledgements}
We would like to thank the members of SFB/TR16 for continuous
encouragement. We acknowledge support from the Deutsche
Forschungsgemeinschaft (DFG) within the SFB/ TR16, and from the
Russian grant RBFF 13-02-00425.

\end{document}